# An Effective Automated Speaking Assessment Approach to Mitigating Data Scarcity and Imbalanced Distribution

**Tien-Hong Lo**[1,2,*]   **Fu-An Chao**[2]   **Tzu-I Wu**[1]   **Yao-Ting Sung**[3]   **Berlin Chen**[1,*]

[1]Department of Computer Science and Information Engineering, National Taiwan Normal University
[2]Research Center for Psychological and Educational Testing, National Taiwan Normal University
[3]Department of Educational Psychology and Counseling, National Taiwan Normal University
`{teinhonglo, fuann, zoetwu, sungtc, berlin}@ntnu.edu.tw`

## Abstract

Automated speaking assessment (ASA) typically involves automatic speech recognition (ASR) and hand-crafted feature extraction from the ASR transcript of a learner's speech. Recently, self-supervised learning (SSL) has shown stellar performance compared to traditional methods. However, SSL-based ASA systems are faced with at least three data-related challenges: limited annotated data, uneven distribution of learner proficiency levels and non-uniform score intervals between different CEFR proficiency levels. To address these challenges, we explore the use of two novel modeling strategies: metric-based classification and loss re-weighting, leveraging distinct SSL-based embedding features. Extensive experimental results on the ICNALE benchmark dataset suggest that our approach can outperform existing strong baselines by a sizable margin, achieving a significant improvement of more than 10% in CEFR prediction accuracy.

## 1 Introduction

With the unprecedented advancements in computer technology and the growing number of second-language (L2) learners worldwide, automated speaking assessment (ASA) has aroused much attention, figuring prominently in computer-assisted language learning (CALL). As shown in Figure 1[1], ASA systems are designed to provide timely feedback on learners' speaking quality, enabling them to improve their spoken language skills in a stress-free and self-directed manner. What is more, ASA systems can alleviate the workload of language teachers and provide a more objective and consistent evaluation on the language proficiency of an L2 learner or test-taker. With the remarkable developments in human language technology, recent years have seen a widespread adoption of ASA systems in CALL, so as to support L2 learners in language acquisition (Moere and Downey, 2016).

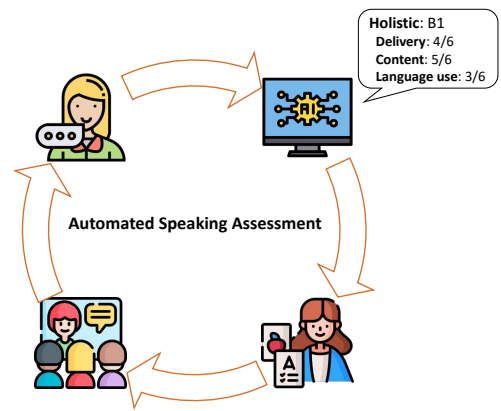

Figure 1: A running example illustrates the scenario of incorporating automated speaking assessment into the traditional classroom.

Iconic ASA approaches involved using standard classifiers and hand-crafted features related to various facets of language proficiency, including but not limited to delivery (such as pronunciation, fluency and intonation), content (such as appropriateness and relevance), and language use (such as vocabulary and grammar) (Strik and Cucchiarini, 1999; Chen et al., 2010; Coutinho et al., 2016; Bhat and Yoon, 2015). In recent years, the rise of self-supervised learning (SSL) paradigms, such as BERT and its variants (Devlin et al., 2019), has opened up new avenues for ASA. These SSL models offer contextualized embeddings that have been successfully integrated into various language assessment tasks like sentence assessment (Arase et al., 2022), grading of essays (Moore et al., 2015; Nadeem et al., 2019; Wu et al., 2023), spoken monologues (Craighead et al., 2020), and many others. On a separate front, the advent of speech-based SSL features has introduced another regime of modeling sophistication and capability to ASA system developments. These features have been particularly effective in specialized tasks within the CALL tasks, such as mispronunciation detection and diagnosis (MDD) (Baevski et al., 2020; Wu et al., 2021; Xu et al., 2021; Peng et al., 2021), automatic pronunciation assessment (APA) (Kim et al., 2022; Chao et al., 2022; Chao et al., 2023) and ASA (Park and Ubale, 2023; McKnight et al., 2023; Banno and Matassoni, 2022; Banno et al., 2023; Li et al., 2023).

---

[*] Corresponding author.

[1] Icons made by Freepik, xnimrodx, and Eucalyp from Flaticon (www.flaticon.com) were used in this paper.

While the work presented in (Banno and Matassoni, 2022; Banno et al., 2023) made pioneering attempts to employ SSL features (BERT and wav2vec 2.0) for ASA, it falls short when faced with three critical issues which our study aims to address: 1) relatively small amount of annotated data, 2) the imbalanced distribution of CEFR proficiency levels (Europe, 2001), and 3) the non-uniform score gaps between different CEFR levels (e.g., $B2 - B1 \neq B1 - A2$). To address these challenges, we first utilize text- and speech-based encoders (BERT and wav2vec 2.0) pre-trained on large-scale datasets. On top of this, this paper introduces effective modeling strategies that are underexplored in previous ASA work, including metric-based classification (Vinyals et al., 2016; Ye and Ling, 2019; Snell et al., 2017; Sun et al., 2019) and loss re-weighting (Conneau and Lample, 2019). In particular, we draw on a unique set of prototypical embeddings for each CEFR level and use various similarity functions to mitigate the imbalanced distribution. It is our hypothesis that metric-based learning not only addresses data imbalance but also efficiently tackles the non-uniform score gaps between CEFR levels.

All variants of our approach are evaluated on the ICNALE corpus (Ishikawa, 2011), an open-source L2 English benchmark dataset, using both text- and speech-based classifiers. Empirical results indicate that our approach can effectively mitigate the issue of data imbalance and achieve significant improvements in accuracy, rising from 77.88% to 92.63% compared to the state-of-the-art baselines (Banno and Matassoni, 2022). Finally, we also conduct a series of analytical experiments to look into the impacts of our modeling strategies on ASA performance, highlighting their practical potential for assessing learners' proficiency. This paper has three-fold contributions:

1. We explore novel and effective modeling strategies for SSL feature extraction and classification in ASA
2. We demonstrate through experiments on the ICNALE corpus that our best-performing instantiation establishes a new state-of-the-art for ASA on this corpus.
3. In particular, our work contributes to the advancement of ASA techniques by addressing challenges related to limited data and imbalanced CEFR-level distribution.

## 2 Related Work

In general, ASA is deployed for assessing speaking proficiency with respect to the responses from an L2 learner, predicting the corresponding level of overall proficiency (holistic score) or specific aspects of proficiency (analytic scores).

It is currently common practice to treat ASA as a classification problem using either text-based or speech-based classifiers. Nevertheless, in the early days, researchers used to tackle the ASA problem with standard classifiers in conjunction with hand-crafted features pertinent to specific facets of language proficiency, such as pronunciation, fluency, prosody, grammar, and others. These features are extracted from the utterances and associated transcripts of an L2 learner and taken as input to a meticulously selected classifier to predict analytic scores (Strik and Cucchiarini, 1999; Chen et al., 2010; Bhat and Yoon, 2015; Moore et al., 2015; Coutinho et al., 2016). For example, Chen et al. (2010) utilized vowel space characteristics for ASA. Bhat and Yoon (2015) ventured into syntactic analysis by employing part-of-speech tag-based complexity measures. Moore et al. (2015) scrutinized the efficacy of the Redshift parser in processing non-native spoken English, finding proficiency in discerning grammatical relations but limitations in detecting speech disfluencies. Coutinho et al. (2016) also concentrated on assessing prosodic and spectral features. Nonetheless, Muller et al. (2009) suggest that hand-crafted features may not always effectively capture important information about proficiency, as their efficacy heavily depends on the underlying assumptions of feature curation.

|       | A2  | B1_1 | B1_2 | B2  | native |
|-------|-----|------|------|-----|--------|
| Train | 299 | 792  | 1681 | 586 | 540    |
| Valid | 16  | 44   | 94   | 33  | 30     |
| Test  | 17  | 44   | 93   | 33  | 30     |

Table 1: Statistical information for each CEFR proficiency level in ICNALE.

As for the feature representations of text-based classification, the NLP community has witnessed a significant trend of transitioning from using static token (e.g., word, subword, and others) embeddings to contextualized word embeddings, such as those derived by BERT (Devlin et al., 2019), with SSL paradigms. Consequently, there has been a surge of research on adopting these contextualized embeddings into automated assessments, such as essays (Nadeem et al., 2019; Arase et al., 2022; Wu et al., 2023) and spoken monologues (Craighead et al., 2020). Nadeem et al. (2019) initiated the use of contextualized embeddings for essay grading. Arase et al. (2022) applied these embeddings in conjunction with prototypical embedding for readability assessment. Craighead et al. (2020) innovatively employed them in evaluating spoken monologues, highlighting the embeddings' versatility in linguistic assessments.

On the other hand, the use of speech-based SSL features emerges as a promising approach. Recent studies have shown good promise in various downstream tasks, such as ASR, and speaker identification (Baevski et al., 2020). Moreover, contextualized representations derived from pre-trained models can capture a diverse range of acoustic and linguistic information for L1 and L2 speech (Shah et al., 2022). This finding adds another dimension to the potential of SSL-based ASA systems. With the increasing availability of annotated speech data and advancements in SSL techniques, there is immense scope for further research and development in the speech process. Despite the promising use of speech-based SSL features in various CALL tasks such as mispronunciation detection and diagnosis (MDD) (Wu et al., 2021; Xu et al., 2021; Peng et al., 2021) and automatic pronunciation assessment (APA) (Kim et al., 2022; Chao et al., 2022; Chao et al., 2023), there is still a dearth of research specifically focused on their application in automated speaking assessment (ASA) (Park and Ubale, 2023; McKnight et al., 2023; Banno and Matassoni, 2022; Banno et al., 2023; Li et al., 2023). This situation presents significant research space and ample opportunity for further exploration and investigation.

## 3 Dataset

To evaluate our proposed approach and corresponding methods, we employed the International Corpus Network of Asian Learners of English (ICNALE) corpus (Ishikawa, 2011), which is a publicly available dataset consisting of written and spoken responses from both native speakers and Asian learners from Japan, China, Hong Kong, South Korea, Taiwan, Singapore, Indonesia, Pakistan, Philippines and Thailand, at various CEFR (Common European Framework of Reference for Language) levels ranging from A2 to B2. Prior to data collection, the ICNALE team assigned CEFR levels to the learners based on their L2 vocabulary size and proficiency scores in English proficiency tests such as IELTS and TOEFL. For our experiments, we made exclusive use of the monologue section of the corpus, consisting of 4,332 speaking responses. In this setup, learners were prompted to describe their opinions on smoking in restaurants and the importance of engaging in part-time employment. Following the methodological practice adopted by (Banno and Matassoni, 2022), this curated collection was divided into a training set of 3,898 responses, as well as a validation set and a test set of 217 responses for each. To evaluate the proficiency of L2 speakers, we will frame the ASA task as a classification problem with five proficiency levels, as illustrated in Table 1.

## 4 Methodology

CEFR levels generally follow an ordinal scale where, for example, the B1 level is considered lower than the B2 level. While it may seem reasonable to approach the ASA task as a regression problem, the non-uniform gaps between the proficiency levels may lead to challenges in interpreting regression outputs (Heilman et al., 2008). As such, we adopt a classification regime to design and implement assessment methods for CEFR-oriented ASA.

A bit of terminology: a training set with $N$ samples $\{(\mathbf{x}_0, y_0), (\mathbf{x}_1, y_1), \cdots, (\mathbf{x}_{N-1}, y_{N-1})\}$ are given, where $\mathbf{x}_i$ is an utterance embedding extracted from either text- or speech-based encoder; $y_i \in \{0, 1, \ldots, J-1\}$ indicates the index $i$ of the corresponding level ($J = 5$ in the ICNALE corpus).

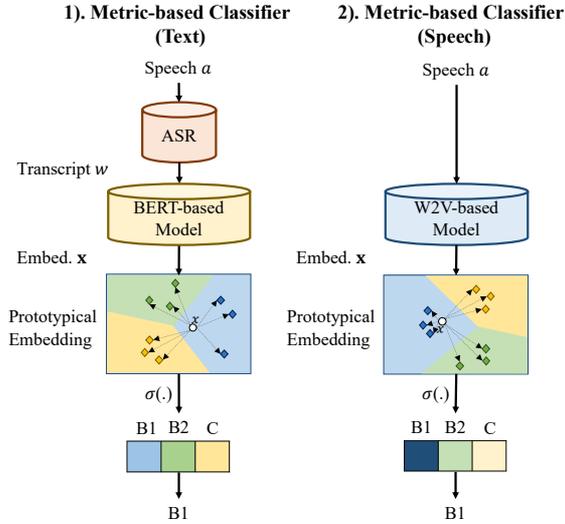

Figure 2: A schematic diagram of proposed models for automated speaking assessment, where σ is the softmax function that aims to choose the maximum value of the prediction vector.

In addition, we utilize two types of SSL-based neural encoders separately as our backbone architectures, namely, a text-based encoder (BERT) and a speech-based encoder (wav2vec 2.0), as schematically depicted in Figure 2. On top of this, we employ two modeling strategies: metric-based classification (*cf.* Section 4.2) and loss re-weighting (*cf.* Section 4.3). These are used to train both text- and speech-based classifiers, with the goal to address the scarcity of learner data at basic (e.g., A1 and A2 speakers) and highly proficient (e.g., C1, C2, and native speakers) levels.

## 4.1 Baseline Classification

### 4.1.1 Text-based classification

This work uses an off-the-shelf BERT architecture to form our text-based encoder $\text{Enc}_{\text{Text}}(*)$. BERT takes an ASR transcript $w_{0:M-1}$ as input, where each word $w_m$ is first transformed into its corresponding token embedding and all words are fed into the encoder layer of BERT altogether to obtain their respective contextualized embedding in a holistic manner:

$$\mathbf{h}_{0:M-1} = \text{TextModel}(w_{0:M-1}) \quad (1)$$

$$\mathbf{x} = \text{MeanPool}(\mathbf{h}_{0:M-1}) \quad (2)$$

where the semantic representation $\mathbf{x} \in \mathbb{R}^d$ is computed by mean pooling of contextualized embedding, $\mathbf{h}_{0:M-1} \in \mathbb{R}^d$, of the transcript. Following that, $\mathbf{x}$ is fed into a multi-layer perceptron (MLP) to predict the corresponding CEFR level. To draw on the benefits of a pre-trained language model, we initialize the model parameters with a pre-trained BERT model and learn the parameterization of the MLP layers from scratch. During training, the entire BERT model is tuned to learn more CEFR-aware knowledge.

As for obtaining the automatic transcripts, we conducted ASR on the ICNALE monologues using the Whisper toolkit (Radford et al., 2022), from which we obtained an average word error rate (WER) of 18.62% with the multi-lingual large model[2] (i.e., the default language is set to English). To avoid losing any possibly existing information about fluency and sentence structure, we managed to retain all relevant information cues, including hesitations, punctuations and others. Notably, since the input to the text-based classifier is an error-prone ASR transcript, it may not accurately represent a learner's proficiency. Moreover, a text-based classifier inevitably fails to capture other important traits of an L2 speaker, such as intonation and prosody.

### 4.1.2 Speech-based classification

We capitalize fully on the off-the-shelf wav2vec 2.0 (W2V) encoder as our speech-based encoder $\text{Enc}_{\text{Speech}}$. W2V is a pre-trained speech model that consists of a feature encoder, a context network, and a quantization module. Analogously to Eq. (1), we encode an input raw waveform $a_{0:T-1}$ with $T$ samples using the last layer of the W2V encoder to obtain latent representation $\mathbf{h}_{0:T_s-1} \in \mathbb{R}^d$ (i.e., $T_s \leq T$):

$$\mathbf{h}_{0:T_s-1} = \text{SpeechModel}(a_{0:T-1}) \quad (3)$$

Then, we employ mean pooling to handle latent representations $\mathbf{h}_{0:T_s-1}$ of different temporal lengths, and the mean-pooled representation is subsequently fed into the MLP layer for classification. Similar to the text-based classifier, we expect that the W2V encoder can produce CEFR-aware representations as well during the training phase.

---
[2] *https://huggingface.co/openai/whisper-large*

## 4.2 Metric-based classification

An imbalanced label distribution (as previously exemplified in Table 1) will lead to overfitting major classes while catering less to minor ones. Because rare classes (referred here to infrequent CEFR levels) must be well considered for real use cases, we alternatively leverage a metric-based classification as a workaround against the label imbalance problem. The metric-based classification has been examined for few-shot learning, where examples are classified based on embedding distances between labeled and unlabeled samples. Previous studies (Vinyals et al., 2016; Ye and Ling, 2019; Snell et al., 2017; Sun et al., 2019) have shown that this type of classifier categorizes samples based on distances within a vector space. The most prevalent techniques in metric-based classification are the matching network (Vinyals et al., 2016; Ye and Ling, 2019) and the prototypical network (Snell et al., 2017; Sun et al., 2019). Bearing some resemblance to an earlier study (Arase et al., 2022) that graded sentence-level text, we adopt a prototypical network to learn embeddings from the training set, which represent CEFR prototypes and predict CEFR levels based on a similarity value. In particular, we extend this approach to investigate speech-based embeddings, disparate similarity functions, and loss re-weighting for enhancing metric-based classification in ASA, which is promising yet underexplored.

After the holistic embeddings of a speaker's spoken response (*cf.* Section 4.1) are obtained, we then adopt the softmax function to compute the distribution $p$ for a response embedding $\mathbf{x}$ over the levels $L_j$ based on similarities to the disparate prototypes:

$$p(y = L_j | \mathbf{x}) = \frac{\exp(s * \text{Sim}(\mathbf{x}, \mathbf{c}_j) + b)}{\sum_J \exp(s * \text{Sim}(\mathbf{x}, \mathbf{c}_j) + b)} \quad (4)$$

where $\text{Sim}(\mathbf{x}, \mathbf{c}_j)$ calculates the similarity between the response embedding $\mathbf{x}$ and a level-specific prototype $\mathbf{c}_j$. Two similarity functions are explored in the prototypical network: one is cosine similarity (dubbed COS) with scaling factors ($s$ and $b$ are learnable parameters) (Chung et al., 2020), and the other is square Euclidean distance (dubbed SED) without scaling factors ($s$ and $b$ are set to 1 and 0, respectively) (Snell et al., 2017):

$$\text{Sim}_{\cos}(\mathbf{x}_i, \mathbf{c}_j) = \frac{\mathbf{x}_i * \mathbf{c}_j}{\|\mathbf{x}_i\| \|\mathbf{c}_j\|} \quad (5)$$

$$\text{Sim}_{\text{sed}}(\mathbf{x}_i, \mathbf{c}_j) = \|\mathbf{x}_i - \mathbf{c}_j\|_2^2 \quad (6)$$

We leverage these two similarity functions in the current work to probe their respective feasibility and performance. When each CEFR level has multiple prototypes ($K > 1$), we compute the mean of the embeddings of these prototypes as the new centroid $\mathbf{c}_j$ or the mean of $K$ similarity values:

$$Sim(\mathbf{x}_i, \mathbf{c}_j) = \frac{1}{K} \sum_k Sim(\mathbf{x}_i, \mathbf{c}_j^k) \quad (7)$$

## 4.3 Loss re-weighting

Apart from the metric-based classifier, we adopt loss re-weighting to address the uneven label distribution, which is based on the multinomial distribution of level frequency and their inverted frequencies (Conneau and Lample, 2019). The loss re-weighting is formulated as:

$$q_i = \frac{p_i^\alpha}{\sum_{j=0}^{J-1} p_j^\alpha} * \frac{1}{p_i} \quad (8)$$

where $p_i$ is to represent the frequency of level $i$ in the training set, and $\alpha \in [0,1]$ regulates the importance weight. A small value of $\alpha$ places a larger weight on infrequent CEFR levels (such as A2 or native). As an aside, we also use a simple loss re-weighting mechanism that only considers the inverted frequencies of CEFR levels in the training set:

$$\hat{q}_i = \frac{\sum_{j=0}^{J-1} p_j}{p_i} \quad (9)$$

The classification loss of all classifiers, including baseline classifiers (*cf.* Section 4.1) and metric-based classifiers (*cf.* Section 4.2), are calculated using cross entropy with (or without) loss re-weighting.

## 5 Experiments

### 5.1 Implementation details

We initialized the model configuration from toolkits provided by HuggingFace (Wolf et al., 2019). The baseline systems are the BERT-based classifier built on *bert-base-uncased* [3] and the wav2vec 2.0-based classifier built on *wav2vec2-base* [4]. The number of prototypes $K$ is set to 3. We use loss re-weighting presented in Eq. (8) for the BERT-based classifier ($\alpha$ is set to 0.5), while using that depicted in Eq. (9) for the wav2vec 2.0-based classifier, both of which are suggested by our preliminary experiments.

---

[3] https://huggingface.co/bert-base-uncased

[4] https://huggingface.co/facebook/wav2vec2-base

| Model | Exp. Tag | RMSE↓ | RMSE$_{MC}$↓ | PCC↑ | ACC↑ | | ACC$_{MC}$↑ | |
|---|---|---|---|---|---|---|---|---|
| | | | | | - | ADJ | - | ADJ |
| BERT* | - | - | - | - | 53.45 | - | - | - |
| W2V* | - | - | - | - | 77.88 | - | - | - |
| BERT | - | 0.948 | 1.028 | 0.628 | 57.60 | 88.02 | 55.53 | 82.99 |
| | + LW | 0.851 | 0.935 | 0.678 | **62.67** | 89.86 | 55.55 | 83.52 |
| | PT(COS) | 0.931 | 1.024 | 0.644 | 58.99 | 89.40 | 52.44 | 84.98 |
| | + LW | 0.877 | 0.937 | 0.674 | 62.21 | 90.78 | **57.89** | 85.21 |
| | PT(SED) | 0.943 | 0.969 | 0.684 | 57.14 | 88.48 | 56.70 | 86.06 |
| | + LW | **0.823** | **0.892** | **0.711** | 59.91 | **93.09** | 53.51 | **88.84** |
| W2V | - | 0.580 | 0.595 | 0.860 | 79.72 | 96.31 | 72.73 | 91.73 |
| | + LW | 0.560 | 0.522 | 0.873 | 75.58 | 97.70 | 75.06 | 96.70 |
| | PT(COS) | 0.539 | 0.605 | 0.876 | 83.41 | 95.85 | 78.03 | 91.06 |
| | + LW | 0.517 | 0.510 | 0.890 | 80.18 | 97.70 | 78.48 | 94.69 |
| | PT(SED) | **0.390** | **0.392** | **0.937** | **92.63** | 98.16 | **90.65** | 96.82 |
| | + LW | 0.429 | 0.426 | 0.924 | 89.40 | **98.16** | 88.51 | 96.59 |

Table 2: Overall performance of our proposed approaches on the ICNALE corpus. * means the results adopted from previous work (Banno and Matassoni, 2022).

| Exp. Tag | ACC↑ | | ACC$_{MC}$↑ | |
|---|---|---|---|---|
| | - | ADJ | - | ADJ |
| PT(COS) | 83.41 | 95.85 | 78.03 | 91.06 |
| PT(SED) | **92.63** | **98.16** | **90.65** | **96.82** |
| PT(COS)$^+$ | 86.64 | 96.31 | 83.53 | 93.02 |
| PT(COS)$^{+*}$ | **89.40** | **96.77** | **84.76** | **93.48** |
| PT(SED)$^+$ | 82.95 | 93.31 | 76.97 | 93.56 |
| PT(SED)$^{+*}$ | 86.63 | 95.85 | 82.23 | 92.26 |

Table 3: Effectiveness of initialization for the prototypical model. The last four rows with the symbol + denotes encoder weight initialization from the vanilla W2V classifier (9th row in Table 2). The symbol * indicates prototypical embedding weight initialization using wav2vec 2.0.

All models were trained on an NVIDIA 3090 GPU using AdamW (Loshchilov and Hutter, 2019) optimizer, with a batch size of 8 and an initial learning rate of 5e-5. The training process of the BERT-based classifier was stopped early with 10 patience epochs based on the averaged macro-accuracy score from the validation set. The training process of the wav2vec 2.0-based classifier was stopped at 10 epochs based on the averaged macro-accuracy score measured on the validation set.

### 5.2 Evaluation metrics

Evaluations of classifiers' effectiveness are crucial for grading applications, where accurate prediction of all levels is essential. However, as the distribution of CEFR levels is unbalanced, conventional evaluation metrics such as accuracy (ACC) and adjacent accuracy (ADJ) may need to be revised. Therefore, macro-type evaluation metrics, ACC$_{MC}$ and RMSE$_{MC}$, were used to penalize models that treats the minor classes poorly. Moreover, since CEFR levels are ordinal, additional evaluation metrics, including root mean squared error (RMSE) and Pearson correlation coefficient (PCC), were employed to evaluate the model performance. These metrics altogether provide a comprehensive evaluation of the classifier's ability to predict all CEFR levels accurately, including minor ones.

### 5.3 Overall performance

At the outset, we report on the performance of two strong baselines, which are SSL-based classifiers on the ICNALE dataset, viz. BERT- and wav2vec 2.0 (W2V)-based classifiers. After that, the prototypical network (PT) in conjunction with loss re-weighting (LW), where the similarity function is either in the form of squared Euclidean distance (SED) or cosine similarity (COS), is employed to enhance the baseline SSL-based classifiers.

In the first set of experiments, we discuss the overall performance of our various methods in comparison with state-of-the-art baselines (Banno and Matassoni, 2022). Table 2 displays the evaluation results of the two classifiers (BERT–and W2V-based baselines) on the ICNALE test set in terms of accuracy and various metrics. W2V exhibits superior performance over BERT in terms of all metrics. Two modeling strategies explored in this study, including loss re-weighting (LW) and prototypical network (PT-COS and PT-SED), achieve superior results than vanilla BERT and W2V across most evaluation metrics, especially for macro-type evaluation metrics (e.g., $RMSE_{MC}$ and $ACC_{MC}$). The experiment results clearly demonstrate the effectiveness of our proposed modeling strategies in enhancing the SSL-based models. Strikingly, with the best setup, which utilizes the W2V-based classifier in conjunction with the prototypical network (SED) and loss re-weighting (LW), our modeling strategy can yield a remarkable improvement in accuracy from 77.88% to 92.63% compared with the start-of-the-art baselines (Banno and Matassoni, 2022). This significant boost in performance confirms the promising potential of our proposed modeling strategies in automated speaking assessment.

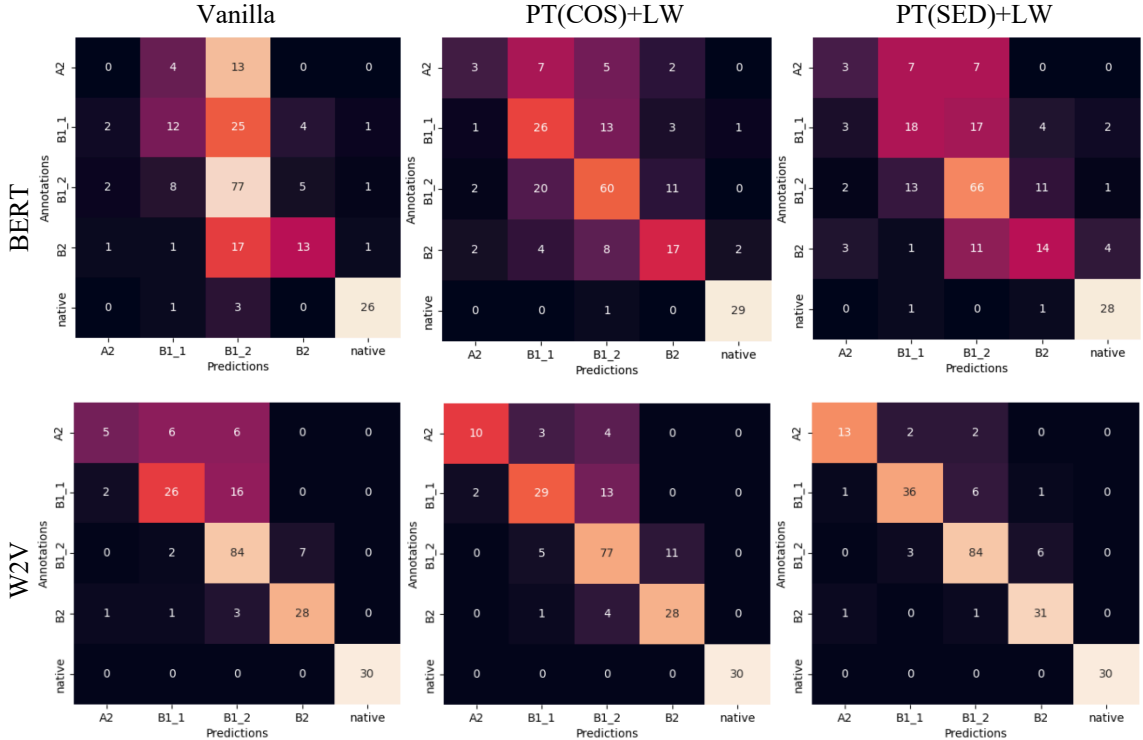

Figure 2: Confusion matrices for SSL-based classifiers using different strategies in predicting proficiency levels.

While Table 2 demonstrates encouraging performance, it's important to note that the effectiveness of the W2V prototypical model, particularly its clustering effect, can vary depending on the dataset's characteristics and the complexity of the speaking tasks. Therefore, further experiments and evaluations are necessary to validate the proposed approach's robustness and generalizability. In the following subsections, we analyze the performance of different models using various evaluation metrics and discuss potential areas for future research and improvement.

### 5.4 Effect of initialization

Table 3 presents the impact of the wav2vec 2.0 encoder on training the metric-based method. Since cosine similarity (referred to as PT(COS)) performs worse than squared Euclidean distance (referred to as PT(SED)), as shown in Table 2, we specifically focus on examining the influence of pretrained embeddings on the performance of PT(COS). From Table 3, it is evident that proper initialization of the encoder and prototypical embeddings have a great impact on the efficacy of the training approach based on the cosine similarity, while it leads to relatively inferior results when using the squared Euclidean distance. Specifically, notable performance improvements are observed by initializing the encoder weights with W2V from the vanilla W2V classifier and using W2V for the weight initialization of prototypical embeddings. However, the results were less pronounced when training with a similarity function based on SED. Our findings highlight the importance of proper initialization of the encoder and prototypical embeddings, meanwhile validating the effectiveness of the cosine similarity in metric-based approaches.

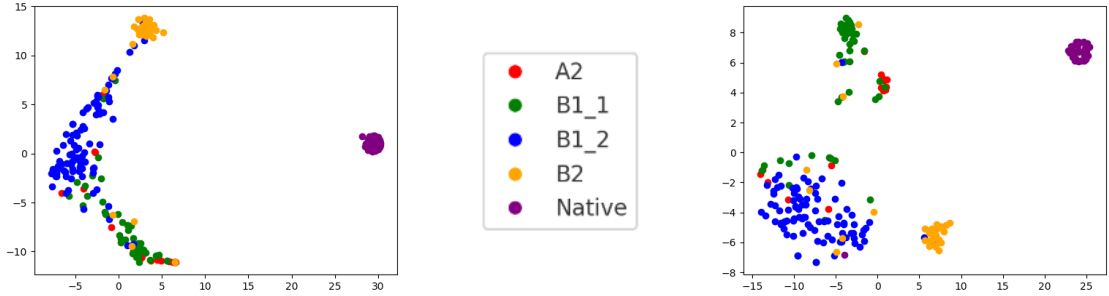

Figure 5: Visualization of CEFR-aware embeddings of vanilla wav2vec2.0 (left) and prototypical wav2vec2.0 with cosine similarity (right). The colors red, green, blue, orange, and purple correspond to A2, B1_1, B1_2, B2, and native speakers, respectively.

### 5.5 Confusion matrices

Figure 3 illustrates the confusion matrices for each CEFR level, showcasing the performance of two SSL-based methods: BERT (top of Figure 3) and W2V (bottom of Figure 3), with the utilization of squared Euclidean distance (SED), cosine similarity (COS) and loss re-weighting (LW). Notably, the W2V-based classifier consistently outperforms the BERT-based classifier across all proficiency levels, demonstrating substantial advancements, particularly for A2, B1_2, and native speakers. The key differentiating factor leading to the superior performance of the W2V-based classifier lies in its ability to capture crucial acoustic, prosodic, and linguistic traits that might be overlooked when relying solely on ASR transcripts. This result signifies the fundamental importance of such latent traits in accurately discerning the distinct CEFR proficiency levels.

### 5.6 ASA Performance for learners from different L1s

Figure 4 plots the histogram of classification performance achieved by our proposed approaches across learners with diverse mother-tongue languages. These tongue languages include Taiwan (TWN), Hong Kong (HKG), Japan (JPN), Korea (KOR), Singapore (SIN), China (CHN), Indonesia (IND), Pakistan (PAK), Philippines (PHL) and Thailand (THA), thereby representing a wide range of language backgrounds. As shown in Figure 4, our best-performing model, W2V-PT(SED), achieves an average accuracy of 90% in predicting CEFR proficiency levels across different tongue languages. Notably, learners from Hong Kong (HKG), Singapore (SIN), Philippines (PHL) and native speakers attained a perfect accuracy rate of 100%.

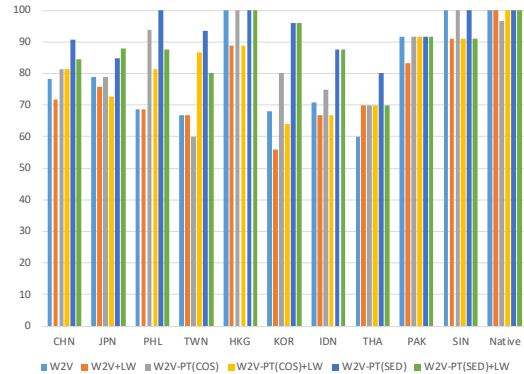

Figure 4: The ASA performance (accuracy%) of our proposed modeling strategies for test learners of different mother-tongue languages.

### 5.7 Visualization of CEFR-aware embeddings

Figure 5 illustrates the t-SNE dimensionality reduction visualization, comparing the original W2V embeddings (left) with the W2V+PT(COS) (right). The left plot shows scattered embeddings forming a manifold. In contrast, the right plot demonstrates an apparent clustering effect, indicating that the W2V prototypical model with cosine similarity successfully groups embeddings by CEFR levels. These results indicate that the metric-based classifier fosters discriminative embeddings reflecting learners' proficiency, which likely account for the improvements in Table 2, underscoring the prototypical approach's efficacy in ASA.

## 6 Conclusion

This paper has put forward two innovative ASA modeling strategies, namely metric-based classification and loss re-weighting, to enhance the performance of self-supervised learning (SSL) models for use in ASA. Both strategies work well with pre-trained embeddings, meanwhile addressing the challenging issues of data scarcity and imbalanced distribution. Extensive experiments on the ICNALE dataset have demonstrated the practical utility of our methods in relation to previous methods in terms of accuracy and various metrics for measuring L2 learners' speaking proficiency. The corresponding results also provide valuable insights for discussing the efficacy of SSL made inroads into ASA. For future work, we plan to delve deeper into the investigation of diverse features and traits, pre-trained models and fine-tuning strategies to mitigate the impact of imbalanced data distribution. Additionally, we envisage extending the scope of our proposed modeling approach to other corpora and tasks, for the purpose of further generalizing its applicability.

## Limitations

The model proposed in this paper focuses on Automated Speaking Assessment using self-supervised learning but exhibits key limitations. Its effectiveness is primarily tied to the ICNALE benchmark dataset, which might not capture the vast diversity of global English learners, potentially limiting the model's generalizability. Additionally, while our model shows promise on this specific dataset, its performance across varied datasets with different learner profiles and proficiency distributions remains untested, raising concerns about its broader applicability. Another challenge is interpreting SSL-based embedding features in relation to specific language proficiency indicators, crucial for enhancing model transparency and facilitating further improvements. To address these issues, we are expanding our dataset to include a design similar to ICNALE, incorporating parallel manual annotation of questions in the corpus. This expansion aims to complement proficiency levels that are currently underrepresented. Furthermore, while we suspect that our method could be applicable to other label-imbalanced classification problems, an empirical investigation of this application is beyond the scope of this paper and is reserved for future research.

## Ethical Considerations

**Bias in Language Assessment** The risk of reinforcing existing educational or linguistic biases is present, especially if the dataset lacks representation from diverse linguistic backgrounds.

**Transparency and Accountability** There's a need for clear communication about how the system assesses language proficiency and mechanisms for feedback to address potential inaccuracies or biases.

**Impact on Learners** If assessments are perceived as unfair, it could affect learners' motivation and confidence, highlighting the importance of aligning the system with diverse learner needs.